\documentclass[aps,prd,twocolumn,showpacs,groupedaddress,nofootinbib]{revtex4-1}

\usepackage{graphicx}
\usepackage{dcolumn}
\usepackage{amsmath}
\usepackage{amssymb}
\usepackage{xcolor}
\usepackage{multirow}
\usepackage{listings}
\usepackage{xcolor}
\usepackage{lineno}
\usepackage{extarrows}
\usepackage[colorlinks,linkcolor=blue,anchorcolor=blue,citecolor=blue]{hyperref}
\allowdisplaybreaks
\newcommand\dx{{\rm d}}
\newcommand\p{\partial}

\begin{document}

\title{Newtonian approximation and possible time-varying $G$ in nonlocal gravities}
\author{S. X. Tian}
\email[]{tshuxun@whu.edu.cn}
\author{Zong-Hong Zhu}
\affiliation{WHU-NAOC Joint Center for Astronomy, School of Physics and Technology, Wuhan University, 430072, Wuhan, China}
\date{\today}
\begin{abstract}
  The Newtonian approximation with a nonvanishing nonlocal background field is analyzed for the scalar-tensor nonlocal gravity and nonlocal Gauss-Bonnet gravity. For these two theories, our calculations show that the Newtonian gravitational constant $G$ is time-varying and $|\dot{G}/G|=\mathcal{O}(H_0)$ for the general case of cosmological background evolution, which is similar to the results of the Deser-Woodard and Maggiore-Mancarella theories. Therefore, observations about the orbit period of binary star (or star-planet) systems could rule out these theories. One thing worth mentioning is that the nonlocal Gauss-Bonnet gravity gives $\Psi=\Phi$ and a constant $G$ in the de Sitter phase. Our results also highlight the uniqueness of the RT model [M. Maggiore, \href{http://dx.doi.org/10.1103/PhysRevD.89.043008}{Phys. Rev. D {\bf 89}, 043008 (2014)}], which is the only nonlocal gravity theory that can successfully describe the gravitational phenomena from solar system to cosmological scales for now.
\end{abstract}
\maketitle

\section{Introduction}\label{sec:01}
Nonlocal gravity is an emergent theory that is used to explain the late-time cosmic acceleration. Deser-Woodard theory \cite{Deser2007}, the first mature nonlocal gravity, adds an $Rf(\Box^{-1}R)$ term to the Einstein-Hilbert action, and can indeed accelerate the late Universe. For this theory, \cite{Koivisto2008a,Elizalde2013} analyzed the dynamics of cosmology, and \cite{Deffayet2009} proposed a method to reconstruct the function $f(x)$ from a known expansion history. The cosmological scalar perturbation analysis shows that Deser-Woodard theory gives a suppressed growth rate for structure formation \cite{Park2013,Nersisyan2017b,Park2018}, which is in better agreement with the redshift-space distortions data than the standard $\Lambda$CDM model. Maggiore-Mancarella theory \cite{Maggiore2014} is also a well-known nonlocal gravity. For this theory, \cite{Barreira2014,Dirian2014,Dirian2016,Maggiore2017,Nersisyan2016,Dirian2017,Belgacem2018,Bellini2018} gave detailed cosmological perturbation analysis, and performed parameter constraints using observational data, including SNIa, BAO, CMB and structure formation data. There are some other nonlocal theories that perform well in cosmology, such as nonlocal Gauss-Bonnet gravity \cite{Capozziello2009,Elizalde2018}, the theory that includes tensor nonlocalities \cite{Barvinsky2012,Ferreira2013,Nersisyan2017a,Tian2018}, nonlocal Galileons \cite{Gabadadze2017}, the RT model of nonlocal gravity \cite{Maggiore2014RT}, the double-metric nonlocal gravity \cite{Vardanyan2018}, and the double nonlocal gravity \cite{Deser2019}. The above results indicate that it is not difficult to explain the late-time acceleration by nonlocal gravity. The remaining issue is whether nonlocal theory can explain the dynamics of gravitational bound systems, e.g., the solar system and the binary neutron star system.

To explain the dynamics of gravitational bound systems, the key is to recover Newtonian gravity or even general relativity in the weak field limit. More precisely, three things are important: The first one is whether the Newtonian gravitational potential $\Phi$ is inversely proportional to $r$ within the range we are considering. Until now, experiments that are used to test the inverse square law of gravitation only give a null result \cite{Hoskins1985,Yang2012,Tan2016}. The second one is whether $\Psi$ is very close to $\Phi$ [see Eq. (\ref{eq:09}) for the meaning of $\Psi$]. The Cassini mission gives $\Psi/\Phi=1+(2.1\pm2.3)\times10^{-5}$ \cite{Bertotti2003}, which is the tightest bound in the solar system today. The third one is whether the Newtonian gravitational constant $G$ keeps constant over cosmic evolution. Time-varying $G$ would influence the periodicity of binary star (or star-planet) systems \cite{Uzan2003,Uzan2011,Shao2016}, and observation gives $|\dot{G}/G|<10^{-12}{\rm yr}^{-1}\approx0.01H_0$ ($95\%$ confidence, $H_0$ is the Hubble constant) or stronger limits based on this idea \cite{Williams2004,Hofmann2010,Fienga2014,Zhu2015,Hofmann2018}. Now we go back to the theoretical things about nonlocal gravity. To analyze the weak field limit, the first step is to choose a background. In general, people may choose the background as the Minkowski metric and set the value of the nonlocal field (e.g., $X\equiv\Box^{-1}R$ in the Deser-Woodard theory) to be zero. Within this framework, \cite{Koivisto2008b}, \cite{Maggiore2014}, and \cite{Tian2018} analyzed the Newtonian approximation for the Deser-Woodard theory, the Maggiore-Mancarella theory, and the scalar-tensor nonlocal gravity, respectively. Similar calculations can be found in the analysis of other nonlocal gravities \cite{Kehagias2014,Conroy2015}. These results are all positive and seem to suggest that obtaining a well-behaved weak field limit is not a tough thing in nonlocal theory. However, the assumption of choosing a vanishing nonlocal field and Minkowski metric as the background is unreasonable. Cosmological evolution will give the nonlocal field a nonzero value, especially when, in the weak field limit case, explicit calculation in a recent paper \cite{Belgacem2019} shows that the cosmological value of the nonlocal field appears in the short scales, and thus the nonlocal field retains its cosmological time dependence down to the short scales. It is important to find the consequences of a nonzero nonlocal field in the Newtonian approximation for various nonlocal gravities. Cosmological scalar perturbation analysis, which takes nonzero nonlocal field as the background, shows that the effective Newtonian gravitational constant $G$ is time-varying in the Deser-Woodard theory \cite{Park2013,Park2018}, and the $|\dot{G}/G|$ constraint rules out this theory. Similar results can be obtained for the Maggiore-Mancarella theory \cite{Barreira2014,Dirian2014}. Do all nonlocal theories suffer from the same problem if we set the background value of the nonlocal field to be nonzero in the weak field limit? In this paper, we discuss this issue for the scalar-tensor nonlocal gravity \cite{Tian2018} and nonlocal Gauss-Bonnet gravity \cite{Capozziello2009}.

This paper is organized as follows: Section \ref{sec:02} summarizes previous works on the weak field limit of the Deser-Woodard and Maggiore-Mancarella theories. Section \ref{sec:03} and Sec. \ref{sec:04} analyze the weak field limit of scalar-tensor nonlocal gravity and nonlocal Gauss-Bonnet gravity, respectively. Our conclusions will be presented in Sec. \ref{sec:05}. One reminder: We develop a Maple program to perform the calculations in this paper as most of the calculations are straightforward but cumbersome, and readers are encouraged to ask the corresponding author for the source code to verify the results.

\section{Newtonian approximation in Deser-Woodard and Maggiore-Mancarella theories}\label{sec:02}
Conventions: The general form of the actions for the gravity theory is $S=\int\dx^4\sqrt{-g}\mathcal{L}_G+S_{\rm m}$, and $\delta S_{\rm m}=-\int\dx^4x\sqrt{-g}T_{\mu\nu}\delta g^{\mu\nu}/2$. The Latin letters run from 1 to 3, and the Greek letters run from 0 to 3.

\subsection{Deser-Woodard theory}
The Deser-Woodard theory is the first mature nonlocal gravity \cite{Deser2007}. The Lagrangian reads
\begin{equation}
  \mathcal{L}_G=\frac{1}{2\kappa}R[1+f(\Box^{-1}R)],
\end{equation}
where $f(\Box^{-1}R)$ is an arbitrary dimensionless function. We can define $X\equiv\Box^{-1}R$ as a pure geometric nonlocal field. Variation of the action with respect to the metric gives the field equations
\begin{align}
  &G_{\mu\nu}+[G_{\mu\nu}+g_{\mu\nu}\Box-\nabla_\mu\nabla_\nu]\{f(X)+\Box^{-1}[Rf'(X)]\}\nonumber\\
  &+[\delta^{(\rho}_\mu\delta^{\sigma)}_{\nu}-\frac{1}{2}g_{\mu\nu}g^{\rho\sigma}]
  [\p_\rho X][\p_\sigma(\Box^{-1}[Rf'(X)])]=\kappa T_{\mu\nu},
\end{align}
where $f'(X)=\dx f(X)/\dx X$. \cite{Deffayet2009} found
\begin{equation}
  f(X)=0.245[\tanh(0.350Y+0.032Y^2+0.003Y^3)-1],
\end{equation}
where $Y=X+16.5$, could give a pretty good expansion history of the Universe.

Adopting $X=0$ and the Minkowski metric for the background, \cite{Koivisto2008b} analyzed the weak field limit of this theory. The result shows observations only constrain the value of the first derivative of $f(X)$. We know that $R=0$ for the Minkowski background, but there is no reason to set $X=\Box^{-1}R=0$, because $\Box^{-1}$ would produce nonzero integral constants. As discussed in \cite{Belgacem2019}, the background value of $X$ used in the weak field limit should be its cosmological scale value. Cosmological scalar perturbation analysis provides such a calculation, and shows that the effective $G$ is time-varying. Combined with $\frac{\dx G}{\dx t}=\frac{\dx G}{\dx z}\frac{\dx z}{\dx t}$ and $\frac{\dx z}{\dx t}|_{\rm today}=-H_0$, which can be obtained by taking the derivative of the cosmological redshift relation $1+z=a_0/a$ with respect to $t$, Fig. 3 in \cite{Park2013} and Fig. 4 in \cite{Park2018} show that $|\dot{G}/G|=\mathcal{O}(H_0)$, and, thus, the Deser-Woodard theory is ruled out if $X$ takes the cosmological scale value in the Newtonian approximation analysis. This conclusion was also confirmed by \cite{Belgacem2019}.

\subsection{Maggiore-Mancarella theory}
The Lagrangian of Maggiore-Mancarella theory \cite{Maggiore2014} is
\begin{equation}
  \mathcal{L}_G=\frac{1}{2\kappa}\left[R-\frac{m^2}{6}R\Box^{-2}R\right],
\end{equation}
where $m$ is constant with a dimension of ${\rm m}^{-1}$. Variation gives the field equations
\begin{align}
  &G_{\mu\nu}-\frac{m^2}{6}\left[-2g_{\mu\nu}U-2\nabla_\mu\nabla_\nu S
  +2SG_{\mu\nu}-(\nabla_\mu S)(\nabla_\nu U)\right.\nonumber\\
  &\left.-(\nabla_\nu S)(\nabla_\mu U)+g_{\mu\nu}(\nabla^\lambda S)(\nabla_\lambda U)-\frac{g_{\mu\nu}}{2}U^2\right]=\kappa T_{\mu\nu},
\end{align}
and $\Box U=-R$, $\Box S=-U$. $U$ and $S$ can be regarded as geometric nonlocal fields. In order to explain the late-time acceleration, we require $m=\mathcal{O}(H_0/c)$ \cite{Maggiore2014}.

\cite{Maggiore2014} also analyzed the Newtonian approximation of this theory with $U,S=0$ in the background, and obtained $\Psi=\Phi\propto1/r$ for $r\ll m^{-1}$. This makes the theory survive from the solar system tests. However, if the nonlocal fields take the cosmological scale value, then the effective Newtonian gravitational constant $G_{\rm eff}=G/(1-m^2\bar{S}/3)$ \cite{Barreira2014}, where bar means cosmological background. The typical cosmological evolution gives $\dot{G}_{\rm eff}/G_{\rm eff}=0.92\times10^{-11}{\rm yr}^{-1}$ \cite{Barreira2014}, which does rule out the Maggiore-Mancarella theory.

In the framework of the Maggiore-Mancarella theory, \cite{Belgacem2019} discussed the issue of how to set the background value of nonlocal fields in the corresponding Newtonian approximation. Even if their analysis is confined in the Maggiore-Mancarella theory, their results should be general for other nonlocal theories. Here we briefly summarize their idea and results. For a nonlocal field $U$ (e.g., $U=-\Box^{-1}R$), the solution should have the form $U_{\rm cosmo}(t)$ at the cosmological scale, and $U_{\rm static}(r)$ at the solar system. The full solution that correctly describes the gravitational phenomena from solar system to cosmological scales should naturally match these two solutions. Using the McVittie-type metric \cite{McVittie1933}, which describes the gravity of a point mass in an expanding Universe, \cite{Belgacem2019} proved the full solution has approximately the form $U_{\rm cosmo}(t)+U_{\rm static}(r)$. \cite{Belgacem2019} also verified this solution with the perturbed Friedmann-Lema\^{i}tre-Robertson-Walker (FLRW) metric. This result means that one cannot ignore the cosmological value of the nonlocal field in the analysis of the Newtonian approximation, and thus the $|\dot{G}/G|$ constraint rules out the Maggiore-Mancarella theory \cite{Belgacem2019}.

\section{Newtonian approximation in scalar-tensor nonlocal gravity}\label{sec:03}
Scalar-tensor nonlocal gravity is used to make $\Psi$ exactly equal to $\Phi$ in the weak field limit \cite{Tian2018}. For the simplest case, the Lagrangian can be written as
\begin{equation}
  \mathcal{L}_G=\frac{1}{2\zeta}\left(R_{\mu\nu}\Box^{-1}R^{\mu\nu}-\frac{1}{2}R\Box^{-1}R\right),
\end{equation}
and the field equations are \cite{Tian2018}
\begin{align}
  &R_{\mu\nu}-g_{\mu\nu}R+\nabla_\mu\nabla_\nu U-UG_{\mu\nu}
  -\frac{1}{2}(\nabla_\mu U)(\nabla_\nu U)\nonumber\\
  &+\frac{g_{\mu\nu}}{4}(\nabla^\lambda U)(\nabla_\lambda U)
  +g_{\mu\nu}\nabla_\alpha\nabla_\beta U^{\alpha\beta}
  -\nabla_\mu\nabla_\alpha U^{\alpha}_{\ \nu}\nonumber\\
  &-\nabla_\nu\nabla_\alpha U^{\alpha}_{\ \mu}
  +2U^{\alpha\beta}R_{\mu\alpha\nu\beta}
  -\frac{g_{\mu\nu}}{4}\Box(U_{\alpha\beta}U^{\alpha\beta})\nonumber\\
  &-\frac{g_{\mu\nu}}{2}R_{\alpha\beta}U^{\alpha\beta}
  +(\nabla_\nu U_{\alpha\beta})(\nabla_\mu U^{\alpha\beta})
  +\nabla_\alpha(U^{\alpha\beta}\nabla_\nu U_{\mu\beta}\nonumber\\
  &+U^{\alpha\beta}\nabla_\mu U_{\nu\beta}
  -U_{\mu\beta}\nabla_\nu U^{\alpha\beta}-U_{\nu\beta}\nabla_\mu U^{\alpha\beta})
  =\zeta T_{\mu\nu},
\end{align}
where the nonlocal field $U_{\mu\nu}\equiv\Box^{-1}R_{\mu\nu}$ (the localized form is $\Box U_{\mu\nu}=R_{\mu\nu}$). The Newtonian approximation with the vanishing nonlocal background field gives $\zeta=8\pi G/c^4$ and the desired Poisson equation \cite{Tian2018}. However, as we discussed before, the nonlocal field should not be equal to zero on the background. In this section, we revisit the weak field limit of this theory with a nonzero nonlocal field.

Because, except for $\zeta$, only dimensionless parameters appear in the Lagrangian, we can still choose the Minkowski metric as the background, and the perturbed metric can be written as
\begin{equation}\label{eq:09}
  \dx s^2=-c^2(1+2\Phi/c^2)\dx t^2+(1-2\Psi/c^2)\dx\mathbf{r}^2,
\end{equation}
where $\Phi=\Phi(\mathbf{r})$ and $\Psi=\Psi(\mathbf{r})$. In the weak field limit, we can assume
\begin{align}
  U_{00}&=f_{10}c^2+f_{11}(\mathbf{r})c^2,\quad U_{0i}=0,\nonumber\\
  U_{ij}&=\delta_{ij}f_{20}+\delta_{ij}f_{21}(\mathbf{r}),
\end{align}
where $f_{i0}$ is the background value and $f_{i1}$ is the first-order perturbation. As shown in \cite{Tian2018}, $f_{i0}$ and $f_{i1}$ are dimensionless. $f_{i0}$ can be regarded as constant in the weak field limit, while it depends on time with cosmic evolution. This is reasonable as $\p/(c\p t)\ll\p/\p x$ (namely quasistatic approximation; see Eq. (3.50) in \cite{Belgacem2019} for an example) in the Newtonian approximation. For the energy-momentum tensor, the only nonzero component is $T_{00}=\rho c^4$.

The above perturbations automatically satisfy the $0i$ component of $\Box U_{\mu\nu}=R_{\mu\nu}$. The $ij$ $(i\neq j)$ component of $\Box U_{\mu\nu}=R_{\mu\nu}$ gives
\begin{equation}
  \frac{\p^2}{\p x^i\p x^j}(\Phi-\Psi)=0.
\end{equation}
Thus we require $\Psi=\Phi$. Combining $\Psi=\Phi$ with the $00$ and $ii$ components of $\Box U_{\mu\nu}=R_{\mu\nu}$ gives
\begin{gather}
  \nabla^2f_{11}-\frac{1}{c^2}(2f_{10}+1)\nabla^2\Phi=0,\\
  \nabla^2f_{21}+\frac{1}{c^2}(2f_{20}-1)\nabla^2\Phi=0,
\end{gather}
respectively. Combining this result with the boundary condition at infinity \cite{Clifton2012,Tian2018}, we obtain
\begin{gather}
  f_{11}=\frac{1}{c^2}(1+2f_{10})\Phi,\\
  f_{21}=\frac{1}{c^2}(1-2f_{20})\Phi.
\end{gather}
Substituting the above results into the field equations, one can find that the $i\mu$ component is automatically established. The $00$ component of the field equations gives
\begin{align}\label{eq:16}
  2[-(f_{10}+f_{20})^2+f_{10}-f_{20}+1]\nabla^2\Phi=\zeta\rho c^4.
\end{align}
Note that $\zeta$ is a constant introduced in the Lagrangian. Compared with the Poisson equation, Eq. (\ref{eq:16}) gives the effective Newtonian gravitational constant
\begin{equation}
  G_{\rm eff}=\frac{\zeta c^4}{8\pi[-(f_{10}+f_{20})^2+f_{10}-f_{20}+1]},
\end{equation}
i.e., $G_{\rm eff}$ is time-varying with cosmic evolution. To explain the late-time evolution, we expect $\dot{f}_{i0}=\mathcal{O}(H_0)$ \cite{Tian2018}, which means $|\dot{G}_{\rm eff}/G_{\rm eff}|=\mathcal{O}(H_0)$. Thus this theory is ruled out by observations.

\section{Newtonian approximation in nonlocal Gauss-Bonnet gravity}\label{sec:04}
For the nonlocal Gauss-Bonnet gravity \cite{Capozziello2009}, one simple case of the Lagrangian is
\begin{equation}\label{eq:18}
  \mathcal{L}_G=\frac{1}{2\kappa}(R-\alpha\mathcal{G}\Box^{-1}\mathcal{G}),
\end{equation}
where $\mathcal{G}=R^2-4R_{\mu\nu}R^{\mu\nu}+R_{\mu\nu\rho\sigma}R^{\mu\nu\rho\sigma}$, and $\alpha$ is constant with a dimension of ${\rm m}^4$. Variation of the action with respect to the metric gives \footnote{Field equations are not given in \cite{Capozziello2009}. We derive this equation based on the standard Lagrange multiplier approach that has been widely used in nonlocal gravity theories.}
\begin{align}\label{eq:19}
  &G_{\mu\nu}-\alpha\left[-8\mathcal{G}G_{\mu\nu}
  +4URR_{\mu\nu}-8UR_{\alpha\nu}R_\mu^{\ \alpha}-g_{\mu\nu}U\mathcal{G}\right.\nonumber\\
  &+(\nabla_\mu U)(\nabla_\nu U)-\frac{g_{\mu\nu}}{2}(\nabla^\alpha U)(\nabla_\alpha U)
  -4R\nabla_\mu\nabla_\nu U\nonumber\\
  &+8R^\alpha_{\ \nu}\nabla_\mu\nabla_\alpha U
  +8R^\alpha_{\ \mu}\nabla_\nu\nabla_\alpha U
  -8g_{\mu\nu}R^{\alpha\beta}\nabla_\alpha\nabla_\beta U\nonumber\\
  &\left.-8R_{\mu\ \ \nu}^{\ \alpha\beta}\nabla_\alpha\nabla_\beta U
  -8UR_{\mu\alpha\nu\beta}R^{\alpha\beta}
  +4UR_{\alpha\beta\mu\gamma}R^{\alpha\beta\ \gamma}_{\ \ \nu}\right]\nonumber\\
  &=\kappa T_{\mu\nu},
\end{align}
where $U\equiv\Box^{-1}\mathcal{G}$ (the localized form is $\Box U=\mathcal{G}$). In order to explain the late-time acceleration, we expect that $\alpha=\mathcal{O}(c^4/H_0^4)$. The dimension and value of $\alpha$ ($\sqrt[4]{\alpha}$ is much bigger than the solar system scale) makes us have to choose the FLRW metric as the background to analyze the Newtonian approximation. For simplicity, we choose the flat FLRW metric. Note that, in general relativity, the linearized field equation, which corresponds to the Poisson equation, is $\frac{1}{a^2}\nabla^2\Phi=4\pi G\rho$ when the background is the flat FLRW metric. The factor $1/a^2$ must be considered when defining the effective Newtonian gravitational constant.

For this theory, \cite{Capozziello2009} discussed the evolution of the Universe. Here we summarize the main results. Substituting the flat FLRW metric into the field equations, we obtain
\begin{gather}
  3H^2-\alpha\left(\frac{1}{2}\dot{U}^2+\frac{24}{c^2}H^3\dot{U}\right)=\kappa\rho c^4,\label{eq:20}\\
  \ddot{U}+3H\dot{U}+\frac{24}{c^2}\frac{\ddot{a}}{a}H^2=0\label{eq:21},
\end{gather}
where $H\equiv\dot{a}/a$. Integrating Eq. (\ref{eq:21}) gives
\begin{equation}\label{eq:22}
  \dot{U}+\frac{8H^3}{c^2}+\frac{C_1}{a^3c^2}=0,
\end{equation}
where $C_1$ is the integral constant. In order to explain the late-time acceleration, we further assume that $C_1/a^3=\mathcal{O}(H^3)$ today, which will be useful in the Newtonian approximation analysis. Then, eliminating $\dot{U}$ in Eq. (\ref{eq:20}), we obtain
\begin{equation}\label{eq:23}
  3H^2+\alpha\left(\frac{160H^6}{c^4}+\frac{16C_1H^3}{a^3c^4}-\frac{C_1^2}{2a^6c^4}\right)=\kappa\rho c^4.
\end{equation}
For $\kappa<0$, $\alpha<0$, and $C_1=0$, Eq. (\ref{eq:23}) gives the scale factor $a\propto t^{3/2}$ in the radiation-dominated era, $a\propto t^2$ in the matter-dominated era, and $a\propto\exp(H_{\rm de}t)$ after the matter-dominated era, where $H_{\rm de}=c[-3/(160\alpha)]^{1/4}$. Eq. (\ref{eq:23}) also shows that the de Sitter Universe [i.e., $a\propto\exp(H_{\rm de}t)$] is a stable attractor regardless of the value of $C_1$.

To analyze the Newtonian approximation, we assume the Universe is filled with pressureless fluid. The perturbed metric can be written as
\begin{equation}\label{eq:24}
  \dx s^2=-c^2(1+2\Phi/c^2)\dx t^2+a^2(1-2\Psi/c^2)\dx\mathbf{r}^2,
\end{equation}
where $a=a(t)$, $\Phi=\Phi(\mathbf{r},t)$, and $\Psi=\Psi(\mathbf{r},t)$. In this analysis, we can assume that $H_0/c\approx\p/(c\p t)\ll\p/(a\p x)$. This assumption and dimensional analysis suggest that the Yukawa-type potential will not appear in our results because $\exp(-r/l_c)\approx\exp(-l_s/l_c)\approx1$, where $l_c$ is the characteristic cosmological scale and $l_s$ is the characteristic solar system scale. Note that, one should be careful when omitting some items in an equation because the value of $\alpha=\mathcal{O}(c^4/H_0^4)$ is much bigger than $l_s^4$. It is a safe way to write down all the items and find the leading part. Similar to \cite{Belgacem2019}, the nonlocal field can be written as
\begin{equation}\label{eq:25}
  U=U_0(t)+U_1(\mathbf{r},t),
\end{equation}
where $U_0$ is the background value and $U_1$ is the first-order perturbation. Based on dimensional analysis, and in order to explain the late-time acceleration, we assume $U_0=\mathcal{O}(H_0^2/c^2)$ today, which is consistent with Eq. (\ref{eq:22}), which says $\dot{U}$ and $8H^3/c^2$ should be of the same order of magnitude in the late-time Universe. In principle, as in general relativity, the nonzero energy-momentum tensors are $T_{00}=\bar{\rho}c^4+\delta\rho\cdot c^4+2\bar{\rho}c^2\Phi$ and $T_{0i}=-\bar{\rho}c^2a^2v^i$. $\bar{\rho}c^2\Phi$ and $\bar{\rho}c^2a^2v^i$ are not the leading terms in Newtonian gravity, so, at the first-order approximation, the only nonzero component is $\delta T_{00}=\delta\rho\cdot c^4$.

Now, we can linearize the field equations with the above perturbations. Substituting Eq. (\ref{eq:24}) and Eq. (\ref{eq:25}) into Eq. (\ref{eq:19}), and considering Eq. (\ref{eq:21}), the $ij$ $(i\neq j)$ component gives
\begin{align}
  &\frac{1}{c^2}\frac{\p^2}{\p x^i\p x^j}\left(\frac{8\alpha\ddot{a}}{a}U_1+\Psi-\Phi
  +\frac{24\alpha H\dot{U}_0}{c^2}\Psi\right.\nonumber\\
  &\qquad\qquad\quad\left.+\frac{8\alpha H\dot{U}_0}{c^2}\Phi
  +\frac{192\alpha H^2\ddot{a}}{ac^4}\Psi\right)=0.
\end{align}
Considering the boundary condition at infinity \cite{Clifton2012,Tian2018}, the above equation gives
\begin{align}\label{eq:27}
  U_1&=-\frac{a}{8\alpha\ddot{a}}\left(\frac{24\alpha H\dot{U}_0}{c^2}\Psi
  +\frac{8\alpha H\dot{U}_0}{c^2}\Phi+\frac{192\alpha H^2\ddot{a}}{ac^4}\Psi\right.\nonumber\\
  &\qquad\qquad\quad\left.+\Psi-\Phi\right).
\end{align}
Substituting the above perturbations into Eq. (\ref{eq:19}), one can find that the leading term of the left side of $0i$ component is proportional to $(H/c^2)\nabla\Phi$, which is identical to the result of general relativity. These equations are used to calculate the velocity $v_i$ of the fluid. Therefore, the above perturbations automatically satisfy the $0i$ component of Eq. (\ref{eq:19}). Combining Eq. (\ref{eq:22}) and Eq. (\ref{eq:27}), omitting nonleading terms, with $\Box U=\mathcal{G}$ gives
\begin{align}
  &\left(-\frac{16\ddot{a}}{a^3c^4}-\frac{1}{8\alpha a\ddot{a}}-\frac{24H^2}{a^2c^4}
  +\frac{3C_1H}{a^4\ddot{a}c^4}+\frac{24H^4}{a\ddot{a}c^4}\right)\nabla^2\Psi\nonumber\\
  &+\left(\frac{1}{8\alpha a\ddot{a}}+\frac{8H^2}{a^2c^4}+\frac{C_1H}{a^4\ddot{a}c^4}
  +\frac{8H^4}{a\ddot{a}c^4}\right)\nabla^2\Phi=0.
\end{align}
Combined with boundary condition at infinity, we obtain
\begin{equation}\label{eq:29}
  \gamma\equiv\frac{\Psi}{\Phi}=
  \frac{c^4/(8\alpha)+8H^4+8H^2\ddot{A}+C_1H/a^3}
  {c^4/(8\alpha)-24H^4+24H^2\ddot{A}+16\ddot{A}^2-3C_1H/a^3},
\end{equation}
where $\ddot{A}\equiv\ddot{a}/a$. Equation (\ref{eq:29}) means generally that $\Psi\neq\Phi$. Equation (\ref{eq:29}) also makes the $ii$ component of Eq. (\ref{eq:19}) automatically established. The $00$ component of Eq. (\ref{eq:19}) gives
\begin{equation}
  \frac{r_1}{a^2}\nabla^2\Psi+\frac{r_2}{a^2}\nabla^2\Phi=\kappa c^4\delta\rho,
\end{equation}
where
\begin{align}
  r_1&=\frac{2H^2}{\ddot{A}}+2-\frac{48\alpha C_1H^3}{\ddot{A}a^3c^4}
  +\frac{384\alpha H^2\ddot{A}}{c^4}+\frac{16\alpha C_1H}{a^3c^4}\nonumber\\
  &\quad-\frac{384\alpha H^6}{\ddot{A}c^4}+\frac{512\alpha H^4}{c^4},\\
  r_2&=-\frac{2H^2}{\ddot{A}}-\frac{16\alpha C_1H^3}{\ddot{A}a^3c^4}
  -\frac{128\alpha H^6}{\ddot{A}c^4}-\frac{192\alpha H^4}{c^4}.
\end{align}
Compared with the Poisson equation, we obtain the effective Newtonian gravitational constant
\begin{equation}\label{eq:33}
  G_{\rm eff}=\frac{\kappa c^4}{4\pi(\gamma r_1+r_2)}.
\end{equation}
Thus, for the general case of background evolution, $G_{\rm eff}$ is time-varying and $|\dot{G}_{\rm eff}/G_{\rm eff}|=\mathcal{O}(H_0)$. This combined with Eq. (\ref{eq:29}) allows us to conclude that the nonlocal Gauss-Bonnet gravity with Eq. (\ref{eq:18}) is ruled out by observations (see the observational constraints on $\Psi/\Phi$ and $\dot{G}/G$ introduced in Sec. \ref{sec:01}).

One interesting result is worth mentioning here. For the de Sitter Universe, $H=c[-3/(160\alpha)]^{1/4}$ and $\ddot{A}=H^2$. Equation (\ref{eq:29}) and Eq. (\ref{eq:33}) give $\gamma=1$ and $G_{\rm eff}=-5\kappa c^4/(32\pi)$, respectively. This result not only shows that $G$ is constant, but also requires that $\kappa<0$, which is consistent with the cosmological result. Note that this is important because the sign of $\kappa$ determines whether the force produced by curved spacetime is attraction or repulsion.

\section{Conclusions}\label{sec:05}
In this paper, we discuss the Newtonian approximation issue of nonlocal gravities. After summarizing the results of the Deser-Woodard and Maggiore-Mancarella theories given by others, we revisit the Newtonian approximation of the scalar-tensor nonlocal gravity with the nonvanishing nonlocal field, and for the first time analyze the weak field limit of the nonlocal Gauss-Bonnet gravity. With the nonzero background value of the nonlocal field, the scalar-tensor nonlocal gravity gives $\Psi=\Phi\propto1/r$, but a time-varying $G$ with $|\dot{G}/G|=\mathcal{O}(H_0)$, and nonlocal Gauss-Bonnet gravity gives $\Psi\neq\Phi$ and $|\dot{G}/G|=\mathcal{O}(H_0)$ for the general case of cosmological background evolution. Observations give $|\dot{G}/G|<0.01H_0$, which rules out the above four nonlocal gravities.

However, not all nonlocal gravities give time-varying $G$. Perturbing the de Sitter Universe in nonlocal Gauss-Bonnet gravity gives $\Psi=\Phi$ and constant $G$, which means we can find a nonlocal theory that behaves well in the solar system for some specific cosmological background evolution. More importantly, \cite{Dirian2014,Nesseris2014,Belgacem2019} pointed out that the RT model \cite{Maggiore2014RT} of nonlocal gravity passes all the tests from the solar system (even considering the cosmological evolution of the nonlocal field) to cosmological scales. Our calculations further confirm the conclusion of \cite{Belgacem2019} that the RT model is the only viable nonlocal gravity theory when confronting observations for now. Uniqueness makes the RT model much more charming and worthy of more attention.

\section*{Acknowledgements}
We are grateful to the referees for valuable comments. This work was supported by the National Natural Science Foundation of China under Grant No. 11633001.


\begin{thebibliography}{45}%
\makeatletter
\providecommand \@ifxundefined [1]{%
 \@ifx{#1\undefined}
}%
\providecommand \@ifnum [1]{%
 \ifnum #1\expandafter \@firstoftwo
 \else \expandafter \@secondoftwo
 \fi
}%
\providecommand \@ifx [1]{%
 \ifx #1\expandafter \@firstoftwo
 \else \expandafter \@secondoftwo
 \fi
}%
\providecommand \natexlab [1]{#1}%
\providecommand \enquote  [1]{``#1''}%
\providecommand \bibnamefont  [1]{#1}%
\providecommand \bibfnamefont [1]{#1}%
\providecommand \citenamefont [1]{#1}%
\providecommand \href@noop [0]{\@secondoftwo}%
\providecommand \href [0]{\begingroup \@sanitize@url \@href}%
\providecommand \@href[1]{\@@startlink{#1}\@@href}%
\providecommand \@@href[1]{\endgroup#1\@@endlink}%
\providecommand \@sanitize@url [0]{\catcode `\\12\catcode `\$12\catcode
  `\&12\catcode `\#12\catcode `\^12\catcode `\_12\catcode `\%12\relax}%
\providecommand \@@startlink[1]{}%
\providecommand \@@endlink[0]{}%
\providecommand \url  [0]{\begingroup\@sanitize@url \@url }%
\providecommand \@url [1]{\endgroup\@href {#1}{\urlprefix }}%
\providecommand \urlprefix  [0]{URL }%
\providecommand \Eprint [0]{\href }%
\providecommand \doibase [0]{http://dx.doi.org/}%
\providecommand \selectlanguage [0]{\@gobble}%
\providecommand \bibinfo  [0]{\@secondoftwo}%
\providecommand \bibfield  [0]{\@secondoftwo}%
\providecommand \translation [1]{[#1]}%
\providecommand \BibitemOpen [0]{}%
\providecommand \bibitemStop [0]{}%
\providecommand \bibitemNoStop [0]{.\EOS\space}%
\providecommand \EOS [0]{\spacefactor3000\relax}%
\providecommand \BibitemShut  [1]{\csname bibitem#1\endcsname}%
\let\auto@bib@innerbib\@empty
\bibitem [{\citenamefont {Deser}\ and\ \citenamefont
  {Woodard}(2007)}]{Deser2007}%
  \BibitemOpen
  \bibfield  {author} {\bibinfo {author} {\bibfnamefont {S.}~\bibnamefont
  {Deser}}\ and\ \bibinfo {author} {\bibfnamefont {R.~P.}\ \bibnamefont
  {Woodard}},\ }\href {\doibase 10.1103/PhysRevLett.99.111301} {\bibfield
  {journal} {\bibinfo  {journal} {Phys. Rev. Lett.}\ }\textbf {\bibinfo
  {volume} {99}},\ \bibinfo {pages} {111301} (\bibinfo {year}
  {2007})}\BibitemShut {NoStop}%
\bibitem [{\citenamefont {Koivisto}(2008{\natexlab{a}})}]{Koivisto2008a}%
  \BibitemOpen
  \bibfield  {author} {\bibinfo {author} {\bibfnamefont {T.}~\bibnamefont
  {Koivisto}},\ }\href {\doibase 10.1103/PhysRevD.77.123513} {\bibfield
  {journal} {\bibinfo  {journal} {Phys. Rev. D}\ }\textbf {\bibinfo {volume}
  {77}},\ \bibinfo {pages} {123513} (\bibinfo {year}
  {2008}{\natexlab{a}})}\BibitemShut {NoStop}%
\bibitem [{\citenamefont {Elizalde}\ \emph {et~al.}(2013)\citenamefont
  {Elizalde}, \citenamefont {Pozdeeva}, \citenamefont {Vernov},\ and\
  \citenamefont {li~Zhang}}]{Elizalde2013}%
  \BibitemOpen
  \bibfield  {author} {\bibinfo {author} {\bibfnamefont {E.}~\bibnamefont
  {Elizalde}}, \bibinfo {author} {\bibfnamefont {E.~O.}\ \bibnamefont
  {Pozdeeva}}, \bibinfo {author} {\bibfnamefont {S.~Y.}\ \bibnamefont
  {Vernov}}, \ and\ \bibinfo {author} {\bibfnamefont {Y.}~\bibnamefont
  {li~Zhang}},\ }\href {\doibase 10.1088/1475-7516/2013/07/034} {\bibfield
  {journal} {\bibinfo  {journal} {J. Cosmol. Astropart. Phys.}\ }
  {\bibinfo {volume} {07}} (\bibinfo {year}
  {2013})\ \bibinfo {pages} {034}}\BibitemShut {NoStop}%
\bibitem [{\citenamefont {Deffayet}\ and\ \citenamefont
  {Woodard}(2009)}]{Deffayet2009}%
  \BibitemOpen
  \bibfield  {author} {\bibinfo {author} {\bibfnamefont {C.}~\bibnamefont
  {Deffayet}}\ and\ \bibinfo {author} {\bibfnamefont {R.~P.}\ \bibnamefont
  {Woodard}},\ }\href {\doibase 10.1088/1475-7516/2009/08/023} {\bibfield
  {journal} {\bibinfo  {journal} {J. Cosmol. Astropart. Phys.}\ }
  {\bibinfo {volume} {08}} (\bibinfo {year}
  {2009})\ \bibinfo {pages} {023}}\BibitemShut {NoStop}%
\bibitem [{\citenamefont {Park}\ and\ \citenamefont
  {Dodelson}(2013)}]{Park2013}%
  \BibitemOpen
  \bibfield  {author} {\bibinfo {author} {\bibfnamefont {S.}~\bibnamefont
  {Park}}\ and\ \bibinfo {author} {\bibfnamefont {S.}~\bibnamefont
  {Dodelson}},\ }\href {\doibase 10.1103/PhysRevD.87.024003} {\bibfield
  {journal} {\bibinfo  {journal} {Phys. Rev. D}\ }\textbf {\bibinfo {volume}
  {87}},\ \bibinfo {pages} {024003} (\bibinfo {year} {2013})}\BibitemShut
  {NoStop}%
\bibitem [{\citenamefont {Nersisyan}\ \emph
  {et~al.}(2017{\natexlab{a}})\citenamefont {Nersisyan}, \citenamefont {Cid},\
  and\ \citenamefont {Amendola}}]{Nersisyan2017b}%
  \BibitemOpen
  \bibfield  {author} {\bibinfo {author} {\bibfnamefont {H.}~\bibnamefont
  {Nersisyan}}, \bibinfo {author} {\bibfnamefont {A.~F.}\ \bibnamefont {Cid}},
  \ and\ \bibinfo {author} {\bibfnamefont {L.}~\bibnamefont {Amendola}},\
  }\href {\doibase 10.1088/1475-7516/2017/04/046} {\bibfield  {journal}
  {\bibinfo  {journal} {J. Cosmol. Astropart. Phys.}\ } {\bibinfo
  {volume} {04}} (\bibinfo {year}
  {2017}{\natexlab{a}})\ \bibinfo {pages} {046}}\BibitemShut {NoStop}%
\bibitem [{\citenamefont {Park}(2018)}]{Park2018}%
  \BibitemOpen
  \bibfield  {author} {\bibinfo {author} {\bibfnamefont {S.}~\bibnamefont
  {Park}},\ }\href {\doibase 10.1103/PhysRevD.97.044006} {\bibfield  {journal}
  {\bibinfo  {journal} {Phys. Rev. D}\ }\textbf {\bibinfo {volume} {97}},\
  \bibinfo {pages} {044006} (\bibinfo {year} {2018})}\BibitemShut {NoStop}%
\bibitem [{\citenamefont {Maggiore}\ and\ \citenamefont
  {Mancarella}(2014)}]{Maggiore2014}%
  \BibitemOpen
  \bibfield  {author} {\bibinfo {author} {\bibfnamefont {M.}~\bibnamefont
  {Maggiore}}\ and\ \bibinfo {author} {\bibfnamefont {M.}~\bibnamefont
  {Mancarella}},\ }\href {\doibase 10.1103/PhysRevD.90.023005} {\bibfield
  {journal} {\bibinfo  {journal} {Phys. Rev. D}\ }\textbf {\bibinfo {volume}
  {90}},\ \bibinfo {pages} {023005} (\bibinfo {year} {2014})}\BibitemShut
  {NoStop}%
\bibitem [{\citenamefont {Barreira}\ \emph {et~al.}(2014)\citenamefont
  {Barreira}, \citenamefont {Li}, \citenamefont {Hellwing}, \citenamefont
  {Baugh},\ and\ \citenamefont {Pascoli}}]{Barreira2014}%
  \BibitemOpen
  \bibfield  {author} {\bibinfo {author} {\bibfnamefont {A.}~\bibnamefont
  {Barreira}}, \bibinfo {author} {\bibfnamefont {B.}~\bibnamefont {Li}},
  \bibinfo {author} {\bibfnamefont {W.~A.}\ \bibnamefont {Hellwing}}, \bibinfo
  {author} {\bibfnamefont {C.~M.}\ \bibnamefont {Baugh}}, \ and\ \bibinfo
  {author} {\bibfnamefont {S.}~\bibnamefont {Pascoli}},\ }\href {\doibase
  10.1088/1475-7516/2014/09/031} {\bibfield  {journal} {\bibinfo  {journal} {J.
  Cosmol. Astropart. Phys.}\ } {\bibinfo {volume} {09}} (\bibinfo {year} {2014})\ \bibinfo
  {pages} {031}}\BibitemShut {NoStop}%
\bibitem [{\citenamefont {Dirian}\ \emph {et~al.}(2014)\citenamefont {Dirian},
  \citenamefont {Foffa}, \citenamefont {Khosravi}, \citenamefont {Kunz},\ and\
  \citenamefont {Maggiore}}]{Dirian2014}%
  \BibitemOpen
  \bibfield  {author} {\bibinfo {author} {\bibfnamefont {Y.}~\bibnamefont
  {Dirian}}, \bibinfo {author} {\bibfnamefont {S.}~\bibnamefont {Foffa}},
  \bibinfo {author} {\bibfnamefont {N.}~\bibnamefont {Khosravi}}, \bibinfo
  {author} {\bibfnamefont {M.}~\bibnamefont {Kunz}}, \ and\ \bibinfo {author}
  {\bibfnamefont {M.}~\bibnamefont {Maggiore}},\ }\href {\doibase
  10.1088/1475-7516/2014/06/033} {\bibfield  {journal} {\bibinfo  {journal} {J.
  Cosmol. Astropart. Phys.}\ } {\bibinfo {volume} {06}} (\bibinfo {year} {2014})\ \bibinfo
  {pages} {033}}\BibitemShut {NoStop}%
\bibitem [{\citenamefont {Dirian}\ \emph {et~al.}(2016)\citenamefont {Dirian},
  \citenamefont {Foffa}, \citenamefont {Kunz}, \citenamefont {Maggiore},\ and\
  \citenamefont {Pettorino}}]{Dirian2016}%
  \BibitemOpen
  \bibfield  {author} {\bibinfo {author} {\bibfnamefont {Y.}~\bibnamefont
  {Dirian}}, \bibinfo {author} {\bibfnamefont {S.}~\bibnamefont {Foffa}},
  \bibinfo {author} {\bibfnamefont {M.}~\bibnamefont {Kunz}}, \bibinfo {author}
  {\bibfnamefont {M.}~\bibnamefont {Maggiore}}, \ and\ \bibinfo {author}
  {\bibfnamefont {V.}~\bibnamefont {Pettorino}},\ }\href {\doibase
  10.1088/1475-7516/2016/05/068} {\bibfield  {journal} {\bibinfo  {journal} {J.
  Cosmol. Astropart. Phys.}\ } {\bibinfo {volume} {05}} (\bibinfo {year} {2016})\ \bibinfo
  {pages} {068}}\BibitemShut {NoStop}%
\bibitem [{\citenamefont {Maggiore}(2017)}]{Maggiore2017}%
  \BibitemOpen
  \bibfield  {author} {\bibinfo {author} {\bibfnamefont {M.}~\bibnamefont
  {Maggiore}},\ }\href {\doibase 10.1007/978-3-319-51700-1_16} {\bibfield
  {journal} {\bibinfo  {journal} {Fund. Theor. Phys.}\ }\textbf {\bibinfo
  {volume} {187}},\ \bibinfo {pages} {221} (\bibinfo {year}
  {2017})}\BibitemShut {NoStop}%
\bibitem [{\citenamefont {Nersisyan}\ \emph {et~al.}(2016)\citenamefont
  {Nersisyan}, \citenamefont {Akrami}, \citenamefont {Amendola}, \citenamefont
  {Koivisto},\ and\ \citenamefont {Rubio}}]{Nersisyan2016}%
  \BibitemOpen
  \bibfield  {author} {\bibinfo {author} {\bibfnamefont {H.}~\bibnamefont
  {Nersisyan}}, \bibinfo {author} {\bibfnamefont {Y.}~\bibnamefont {Akrami}},
  \bibinfo {author} {\bibfnamefont {L.}~\bibnamefont {Amendola}}, \bibinfo
  {author} {\bibfnamefont {T.~S.}\ \bibnamefont {Koivisto}}, \ and\ \bibinfo
  {author} {\bibfnamefont {J.}~\bibnamefont {Rubio}},\ }\href {\doibase
  10.1103/PhysRevD.94.043531} {\bibfield  {journal} {\bibinfo  {journal} {Phys.
  Rev. D}\ }\textbf {\bibinfo {volume} {94}},\ \bibinfo {pages} {043531}
  (\bibinfo {year} {2016})}\BibitemShut {NoStop}%
\bibitem [{\citenamefont {Dirian}(2017)}]{Dirian2017}%
  \BibitemOpen
  \bibfield  {author} {\bibinfo {author} {\bibfnamefont {Y.}~\bibnamefont
  {Dirian}},\ }\href {\doibase 10.1103/PhysRevD.96.083513} {\bibfield
  {journal} {\bibinfo  {journal} {Phys. Rev. D}\ }\textbf {\bibinfo {volume}
  {96}},\ \bibinfo {pages} {083513} (\bibinfo {year} {2017})}\BibitemShut
  {NoStop}%
\bibitem [{\citenamefont {Belgacem}\ \emph {et~al.}(2018)\citenamefont
  {Belgacem}, \citenamefont {Dirian}, \citenamefont {Foffa},\ and\
  \citenamefont {Maggiore}}]{Belgacem2018}%
  \BibitemOpen
  \bibfield  {author} {\bibinfo {author} {\bibfnamefont {E.}~\bibnamefont
  {Belgacem}}, \bibinfo {author} {\bibfnamefont {Y.}~\bibnamefont {Dirian}},
  \bibinfo {author} {\bibfnamefont {S.}~\bibnamefont {Foffa}}, \ and\ \bibinfo
  {author} {\bibfnamefont {M.}~\bibnamefont {Maggiore}},\ }\href {\doibase
  10.1088/1475-7516/2018/03/002} {\bibfield  {journal} {\bibinfo  {journal} {J.
  Cosmol. Astropart. Phys.}\ } {\bibinfo {volume} {03}} (\bibinfo {year} {2018})\ \bibinfo
  {pages} {002}}\BibitemShut {NoStop}%
\bibitem [{\citenamefont {Bellini}\ \emph {et~al.}(2018)\citenamefont
  {Bellini}, \citenamefont {Barreira}, \citenamefont {Frusciante},
  \citenamefont {Hu}, \citenamefont {Peirone}, \citenamefont {Raveri},
  \citenamefont {Zumalac\'arregui}, \citenamefont {Avilez-Lopez}, \citenamefont
  {Ballardini}, \citenamefont {Battye}, \citenamefont {Bolliet}, \citenamefont
  {Calabrese}, \citenamefont {Dirian}, \citenamefont {Ferreira}, \citenamefont
  {Finelli}, \citenamefont {Huang}, \citenamefont {Ivanov}, \citenamefont
  {Lesgourgues}, \citenamefont {Li}, \citenamefont {Lima}, \citenamefont
  {Pace}, \citenamefont {Paoletti}, \citenamefont {Sawicki}, \citenamefont
  {Silvestri}, \citenamefont {Skordis}, \citenamefont {Umilt\`a},\ and\
  \citenamefont {Vernizzi}}]{Bellini2018}%
  \BibitemOpen
  \bibfield  {author} {\bibinfo {author} {\bibfnamefont {E.}~\bibnamefont
  {Bellini}}, \bibinfo {author} {\bibfnamefont {A.}~\bibnamefont {Barreira}},
  \bibinfo {author} {\bibfnamefont {N.}~\bibnamefont {Frusciante}}, \bibinfo
  {author} {\bibfnamefont {B.}~\bibnamefont {Hu}}, \bibinfo {author}
  {\bibfnamefont {S.}~\bibnamefont {Peirone}}, \bibinfo {author} {\bibfnamefont
  {M.}~\bibnamefont {Raveri}}, \bibinfo {author} {\bibfnamefont
  {M.}~\bibnamefont {Zumalac\'arregui}}, \bibinfo {author} {\bibfnamefont
  {A.}~\bibnamefont {Avilez-Lopez}}, \bibinfo {author} {\bibfnamefont
  {M.}~\bibnamefont {Ballardini}}, \bibinfo {author} {\bibfnamefont {R.~A.}\
  \bibnamefont {Battye}}, \bibinfo {author} {\bibfnamefont {B.}~\bibnamefont
  {Bolliet}}, \bibinfo {author} {\bibfnamefont {E.}~\bibnamefont {Calabrese}},
  \bibinfo {author} {\bibfnamefont {Y.}~\bibnamefont {Dirian}}, \bibinfo
  {author} {\bibfnamefont {P.~G.}\ \bibnamefont {Ferreira}}, \bibinfo {author}
  {\bibfnamefont {F.}~\bibnamefont {Finelli}}, \bibinfo {author} {\bibfnamefont
  {Z.}~\bibnamefont {Huang}}, \bibinfo {author} {\bibfnamefont {M.~M.}\
  \bibnamefont {Ivanov}}, \bibinfo {author} {\bibfnamefont {J.}~\bibnamefont
  {Lesgourgues}}, \bibinfo {author} {\bibfnamefont {B.}~\bibnamefont {Li}},
  \bibinfo {author} {\bibfnamefont {N.~A.}\ \bibnamefont {Lima}}, \bibinfo
  {author} {\bibfnamefont {F.}~\bibnamefont {Pace}}, \bibinfo {author}
  {\bibfnamefont {D.}~\bibnamefont {Paoletti}}, \bibinfo {author}
  {\bibfnamefont {I.}~\bibnamefont {Sawicki}}, \bibinfo {author} {\bibfnamefont
  {A.}~\bibnamefont {Silvestri}}, \bibinfo {author} {\bibfnamefont
  {C.}~\bibnamefont {Skordis}}, \bibinfo {author} {\bibfnamefont
  {C.}~\bibnamefont {Umilt\`a}}, \ and\ \bibinfo {author} {\bibfnamefont
  {F.}~\bibnamefont {Vernizzi}},\ }\href {\doibase 10.1103/PhysRevD.97.023520}
  {\bibfield  {journal} {\bibinfo  {journal} {Phys. Rev. D}\ }\textbf {\bibinfo
  {volume} {97}},\ \bibinfo {pages} {023520} (\bibinfo {year}
  {2018})}\BibitemShut {NoStop}%
\bibitem [{\citenamefont {Capozziello}\ \emph {et~al.}(2009)\citenamefont
  {Capozziello}, \citenamefont {Elizalde}, \citenamefont {Nojiri},\ and\
  \citenamefont {Odintsov}}]{Capozziello2009}%
  \BibitemOpen
  \bibfield  {author} {\bibinfo {author} {\bibfnamefont {S.}~\bibnamefont
  {Capozziello}}, \bibinfo {author} {\bibfnamefont {E.}~\bibnamefont
  {Elizalde}}, \bibinfo {author} {\bibfnamefont {S.}~\bibnamefont {Nojiri}}, \
  and\ \bibinfo {author} {\bibfnamefont {S.~D.}\ \bibnamefont {Odintsov}},\
  }\href {\doibase 10.1016/j.physletb.2008.11.060} {\bibfield  {journal}
  {\bibinfo  {journal} {Phys. Lett. B}\ }\textbf {\bibinfo {volume} {671}},\
  \bibinfo {pages} {193} (\bibinfo {year} {2009})}\BibitemShut {NoStop}%
\bibitem [{\citenamefont {Elizalde}\ \emph {et~al.}(2018)\citenamefont
  {Elizalde}, \citenamefont {Odintsov}, \citenamefont {Pozdeeva},\ and\
  \citenamefont {Vernov}}]{Elizalde2018}%
  \BibitemOpen
  \bibfield  {author} {\bibinfo {author} {\bibfnamefont {E.}~\bibnamefont
  {Elizalde}}, \bibinfo {author} {\bibfnamefont {S.~D.}\ \bibnamefont
  {Odintsov}}, \bibinfo {author} {\bibfnamefont {E.~O.}\ \bibnamefont
  {Pozdeeva}}, \ and\ \bibinfo {author} {\bibfnamefont {S.~Y.}\ \bibnamefont
  {Vernov}},\ }\href {\doibase 10.1142/S0219887818501888} {\bibfield  {journal}
  {\bibinfo  {journal} {Int. J. Geom. Methods Mod. Phys.}\ }\textbf {\bibinfo
  {volume} {15}},\ \bibinfo {pages} {1850188} (\bibinfo {year}
  {2018})}\BibitemShut {NoStop}%
\bibitem [{\citenamefont {Barvinsky}(2012)}]{Barvinsky2012}%
  \BibitemOpen
  \bibfield  {author} {\bibinfo {author} {\bibfnamefont {A.~O.}\ \bibnamefont
  {Barvinsky}},\ }\href {\doibase 10.1016/j.physletb.2012.02.075} {\bibfield
  {journal} {\bibinfo  {journal} {Phys. Lett. B}\ }\textbf {\bibinfo {volume}
  {710}},\ \bibinfo {pages} {12} (\bibinfo {year} {2012})}\BibitemShut
  {NoStop}%
\bibitem [{\citenamefont {Ferreira}\ and\ \citenamefont
  {Maroto}(2013)}]{Ferreira2013}%
  \BibitemOpen
  \bibfield  {author} {\bibinfo {author} {\bibfnamefont {P.~G.}\ \bibnamefont
  {Ferreira}}\ and\ \bibinfo {author} {\bibfnamefont {A.~L.}\ \bibnamefont
  {Maroto}},\ }\href {\doibase 10.1103/PhysRevD.88.123502} {\bibfield
  {journal} {\bibinfo  {journal} {Phys. Rev. D}\ }\textbf {\bibinfo {volume}
  {88}},\ \bibinfo {pages} {123502} (\bibinfo {year} {2013})}\BibitemShut
  {NoStop}%
\bibitem [{\citenamefont {Nersisyan}\ \emph
  {et~al.}(2017{\natexlab{b}})\citenamefont {Nersisyan}, \citenamefont
  {Akrami}, \citenamefont {Amendola}, \citenamefont {Koivisto}, \citenamefont
  {Rubio},\ and\ \citenamefont {Solomon}}]{Nersisyan2017a}%
  \BibitemOpen
  \bibfield  {author} {\bibinfo {author} {\bibfnamefont {H.}~\bibnamefont
  {Nersisyan}}, \bibinfo {author} {\bibfnamefont {Y.}~\bibnamefont {Akrami}},
  \bibinfo {author} {\bibfnamefont {L.}~\bibnamefont {Amendola}}, \bibinfo
  {author} {\bibfnamefont {T.~S.}\ \bibnamefont {Koivisto}}, \bibinfo {author}
  {\bibfnamefont {J.}~\bibnamefont {Rubio}}, \ and\ \bibinfo {author}
  {\bibfnamefont {A.~R.}\ \bibnamefont {Solomon}},\ }\href {\doibase
  10.1103/PhysRevD.95.043539} {\bibfield  {journal} {\bibinfo  {journal} {Phys.
  Rev. D}\ }\textbf {\bibinfo {volume} {95}},\ \bibinfo {pages} {043539}
  (\bibinfo {year} {2017}{\natexlab{b}})}\BibitemShut {NoStop}%
\bibitem [{\citenamefont {Tian}(2018)}]{Tian2018}%
  \BibitemOpen
  \bibfield  {author} {\bibinfo {author} {\bibfnamefont {S.}~\bibnamefont
  {Tian}},\ }\href {\doibase 10.1103/PhysRevD.98.084040} {\bibfield  {journal}
  {\bibinfo  {journal} {Phys. Rev. D}\ }\textbf {\bibinfo {volume} {98}},\
  \bibinfo {pages} {084040} (\bibinfo {year} {2018})}\BibitemShut {NoStop}%
\bibitem [{\citenamefont {Gabadadze}\ and\ \citenamefont
  {Yu}(2017)}]{Gabadadze2017}%
  \BibitemOpen
  \bibfield  {author} {\bibinfo {author} {\bibfnamefont {G.}~\bibnamefont
  {Gabadadze}}\ and\ \bibinfo {author} {\bibfnamefont {S.}~\bibnamefont {Yu}},\
  }\href {\doibase 10.1016/j.physletb.2017.03.027} {\bibfield  {journal}
  {\bibinfo  {journal} {Phys. Lett. B}\ }\textbf {\bibinfo {volume} {768}},\
  \bibinfo {pages} {397} (\bibinfo {year} {2017})}\BibitemShut {NoStop}%
\bibitem [{\citenamefont {Maggiore}(2014)}]{Maggiore2014RT}%
  \BibitemOpen
  \bibfield  {author} {\bibinfo {author} {\bibfnamefont {M.}~\bibnamefont
  {Maggiore}},\ }\href {\doibase 10.1103/PhysRevD.89.043008} {\bibfield
  {journal} {\bibinfo  {journal} {Phys. Rev. D}\ }\textbf {\bibinfo {volume}
  {89}},\ \bibinfo {pages} {043008} (\bibinfo {year} {2014})}\BibitemShut
  {NoStop}%
\bibitem [{\citenamefont {Vardanyan}\ \emph {et~al.}(2018)\citenamefont
  {Vardanyan}, \citenamefont {Akrami}, \citenamefont {Amendola},\ and\
  \citenamefont {Silvestri}}]{Vardanyan2018}%
  \BibitemOpen
  \bibfield  {author} {\bibinfo {author} {\bibfnamefont {V.}~\bibnamefont
  {Vardanyan}}, \bibinfo {author} {\bibfnamefont {Y.}~\bibnamefont {Akrami}},
  \bibinfo {author} {\bibfnamefont {L.}~\bibnamefont {Amendola}}, \ and\
  \bibinfo {author} {\bibfnamefont {A.}~\bibnamefont {Silvestri}},\ }\href
  {\doibase 10.1088/1475-7516/2018/03/048} {\bibfield  {journal} {\bibinfo
  {journal} {J. Cosmol. Astropart. Phys.}\ } {\bibinfo {volume} {03}} (\bibinfo {year} {2018})\
  \bibinfo {pages} {048}}\BibitemShut {NoStop}%
\bibitem [{\citenamefont {Deser}\ and\ \citenamefont
  {Woodard}(2019)}]{Deser2019}%
  \BibitemOpen
  \bibfield  {author} {\bibinfo {author} {\bibfnamefont {S.}~\bibnamefont
  {Deser}}\ and\ \bibinfo {author} {\bibfnamefont {R.~P.}\ \bibnamefont
  {Woodard}},}\ \Eprint
  {http://arxiv.org/abs/1902.08075} {arXiv:1902.08075} \BibitemShut {NoStop}%
\bibitem [{\citenamefont {Hoskins}\ \emph {et~al.}(1985)\citenamefont
  {Hoskins}, \citenamefont {Newman}, \citenamefont {Spero},\ and\ \citenamefont
  {Schultz}}]{Hoskins1985}%
  \BibitemOpen
  \bibfield  {author} {\bibinfo {author} {\bibfnamefont {J.~K.}\ \bibnamefont
  {Hoskins}}, \bibinfo {author} {\bibfnamefont {R.~D.}\ \bibnamefont {Newman}},
  \bibinfo {author} {\bibfnamefont {R.}~\bibnamefont {Spero}}, \ and\ \bibinfo
  {author} {\bibfnamefont {J.}~\bibnamefont {Schultz}},\ }\href {\doibase
  10.1103/PhysRevD.32.3084} {\bibfield  {journal} {\bibinfo  {journal} {Phys.
  Rev. D}\ }\textbf {\bibinfo {volume} {32}},\ \bibinfo {pages} {3084}
  (\bibinfo {year} {1985})}\BibitemShut {NoStop}%
\bibitem [{\citenamefont {Yang}\ \emph {et~al.}(2012)\citenamefont {Yang},
  \citenamefont {Zhan}, \citenamefont {Wang}, \citenamefont {Shao},
  \citenamefont {Tu}, \citenamefont {Tan},\ and\ \citenamefont
  {Luo}}]{Yang2012}%
  \BibitemOpen
  \bibfield  {author} {\bibinfo {author} {\bibfnamefont {S.-Q.}\ \bibnamefont
  {Yang}}, \bibinfo {author} {\bibfnamefont {B.-F.}\ \bibnamefont {Zhan}},
  \bibinfo {author} {\bibfnamefont {Q.-L.}\ \bibnamefont {Wang}}, \bibinfo
  {author} {\bibfnamefont {C.-G.}\ \bibnamefont {Shao}}, \bibinfo {author}
  {\bibfnamefont {L.-C.}\ \bibnamefont {Tu}}, \bibinfo {author} {\bibfnamefont
  {W.-H.}\ \bibnamefont {Tan}}, \ and\ \bibinfo {author} {\bibfnamefont
  {J.}~\bibnamefont {Luo}},\ }\href {\doibase 10.1103/PhysRevLett.108.081101}
  {\bibfield  {journal} {\bibinfo  {journal} {Phys. Rev. Lett.}\ }\textbf
  {\bibinfo {volume} {108}},\ \bibinfo {pages} {081101} (\bibinfo {year}
  {2012})}\BibitemShut {NoStop}%
\bibitem [{\citenamefont {Tan}\ \emph {et~al.}(2016)\citenamefont {Tan},
  \citenamefont {Yang}, \citenamefont {Shao}, \citenamefont {Li}, \citenamefont
  {Du}, \citenamefont {Zhan}, \citenamefont {Wang}, \citenamefont {Luo},
  \citenamefont {Tu},\ and\ \citenamefont {Luo}}]{Tan2016}%
  \BibitemOpen
  \bibfield  {author} {\bibinfo {author} {\bibfnamefont {W.-H.}\ \bibnamefont
  {Tan}}, \bibinfo {author} {\bibfnamefont {S.-Q.}\ \bibnamefont {Yang}},
  \bibinfo {author} {\bibfnamefont {C.-G.}\ \bibnamefont {Shao}}, \bibinfo
  {author} {\bibfnamefont {J.}~\bibnamefont {Li}}, \bibinfo {author}
  {\bibfnamefont {A.-B.}\ \bibnamefont {Du}}, \bibinfo {author} {\bibfnamefont
  {B.-F.}\ \bibnamefont {Zhan}}, \bibinfo {author} {\bibfnamefont {Q.-L.}\
  \bibnamefont {Wang}}, \bibinfo {author} {\bibfnamefont {P.-S.}\ \bibnamefont
  {Luo}}, \bibinfo {author} {\bibfnamefont {L.-C.}\ \bibnamefont {Tu}}, \ and\
  \bibinfo {author} {\bibfnamefont {J.}~\bibnamefont {Luo}},\ }\href {\doibase
  10.1103/PhysRevLett.116.131101} {\bibfield  {journal} {\bibinfo  {journal}
  {Phys. Rev. Lett.}\ }\textbf {\bibinfo {volume} {116}},\ \bibinfo {pages}
  {131101} (\bibinfo {year} {2016})}\BibitemShut {NoStop}%
\bibitem [{\citenamefont {Bertotti}\ \emph {et~al.}(2003)\citenamefont
  {Bertotti}, \citenamefont {Iess},\ and\ \citenamefont
  {Tortora}}]{Bertotti2003}%
  \BibitemOpen
  \bibfield  {author} {\bibinfo {author} {\bibfnamefont {B.}~\bibnamefont
  {Bertotti}}, \bibinfo {author} {\bibfnamefont {L.}~\bibnamefont {Iess}}, \
  and\ \bibinfo {author} {\bibfnamefont {P.}~\bibnamefont {Tortora}},\ }\href
  {\doibase 10.1038/nature01997} {\bibfield  {journal} {\bibinfo  {journal}
  {Nature (London)}\ }\textbf {\bibinfo {volume} {425}},\ \bibinfo {pages}
  {374} (\bibinfo {year} {2003})}\BibitemShut {NoStop}%
\bibitem [{\citenamefont {Uzan}(2003)}]{Uzan2003}%
  \BibitemOpen
  \bibfield  {author} {\bibinfo {author} {\bibfnamefont {J.-P.}\ \bibnamefont
  {Uzan}},\ }\href {\doibase 10.1103/RevModPhys.75.403} {\bibfield  {journal}
  {\bibinfo  {journal} {Rev. Mod. Phys.}\ }\textbf {\bibinfo {volume} {75}},\
  \bibinfo {pages} {403} (\bibinfo {year} {2003})}\BibitemShut {NoStop}%
\bibitem [{\citenamefont {Uzan}(2011)}]{Uzan2011}%
  \BibitemOpen
  \bibfield  {author} {\bibinfo {author} {\bibfnamefont {J.-P.}\ \bibnamefont
  {Uzan}},\ }\href {\doibase 10.12942/lrr-2011-2} {\bibfield  {journal}
  {\bibinfo  {journal} {Living Rev. Relativity}\ }\textbf {\bibinfo {volume}
  {14}},\ \bibinfo {pages} {2} (\bibinfo {year} {2011})}\BibitemShut {NoStop}%
\bibitem [{\citenamefont {Shao}\ and\ \citenamefont {Wex}(2016)}]{Shao2016}%
  \BibitemOpen
  \bibfield  {author} {\bibinfo {author} {\bibfnamefont {L.}~\bibnamefont
  {Shao}}\ and\ \bibinfo {author} {\bibfnamefont {N.}~\bibnamefont {Wex}},\
  }\href {\doibase 10.1007/s11433-016-0087-6} {\bibfield  {journal} {\bibinfo
  {journal} {Sci. China Phys. Mech. Astron.}\ }\textbf {\bibinfo {volume}
  {59}},\ \bibinfo {pages} {699501} (\bibinfo {year} {2016})}\BibitemShut
  {NoStop}%
\bibitem [{\citenamefont {Williams}\ \emph {et~al.}(2004)\citenamefont
  {Williams}, \citenamefont {Turyshev},\ and\ \citenamefont
  {Boggs}}]{Williams2004}%
  \BibitemOpen
  \bibfield  {author} {\bibinfo {author} {\bibfnamefont {J.~G.}\ \bibnamefont
  {Williams}}, \bibinfo {author} {\bibfnamefont {S.~G.}\ \bibnamefont
  {Turyshev}}, \ and\ \bibinfo {author} {\bibfnamefont {D.~H.}\ \bibnamefont
  {Boggs}},\ }\href {\doibase 10.1103/PhysRevLett.93.261101} {\bibfield
  {journal} {\bibinfo  {journal} {Phys. Rev. Lett.}\ }\textbf {\bibinfo
  {volume} {93}},\ \bibinfo {pages} {261101} (\bibinfo {year}
  {2004})}\BibitemShut {NoStop}%
\bibitem [{\citenamefont {Hofmann}\ \emph {et~al.}(2010)\citenamefont
  {Hofmann}, \citenamefont {M\"uller},\ and\ \citenamefont
  {Biskupek}}]{Hofmann2010}%
  \BibitemOpen
  \bibfield  {author} {\bibinfo {author} {\bibfnamefont {F.}~\bibnamefont
  {Hofmann}}, \bibinfo {author} {\bibfnamefont {J.}~\bibnamefont {M\"uller}}, \
  and\ \bibinfo {author} {\bibfnamefont {L.}~\bibnamefont {Biskupek}},\ }\href
  {\doibase 10.1051/0004-6361/201015659} {\bibfield  {journal} {\bibinfo
  {journal} {Astron. Astrophys.}\ }\textbf {\bibinfo {volume} {522}},\ \bibinfo
  {pages} {L5} (\bibinfo {year} {2010})}\BibitemShut {NoStop}%
\bibitem [{\citenamefont {Fienga}\ \emph {et~al.}(2014)\citenamefont {Fienga},
  \citenamefont {Laskar}, \citenamefont {Exertier}, \citenamefont {Manche},\
  and\ \citenamefont {Gastineau}}]{Fienga2014}%
  \BibitemOpen
  \bibfield  {author} {\bibinfo {author} {\bibfnamefont {A.}~\bibnamefont
  {Fienga}}, \bibinfo {author} {\bibfnamefont {J.}~\bibnamefont {Laskar}},
  \bibinfo {author} {\bibfnamefont {P.}~\bibnamefont {Exertier}}, \bibinfo
  {author} {\bibfnamefont {H.}~\bibnamefont {Manche}}, \ and\ \bibinfo {author}
  {\bibfnamefont {M.}~\bibnamefont {Gastineau}},}\ \Eprint {http://arxiv.org/abs/1409.4932} {arXiv:1409.4932}
  \BibitemShut {NoStop}%
\bibitem [{\citenamefont {Zhu}\ \emph {et~al.}(2015)\citenamefont {Zhu},
  \citenamefont {Stairs}, \citenamefont {Demorest}, \citenamefont {Nice},
  \citenamefont {Ellis}, \citenamefont {Ransom}, \citenamefont {Arzoumanian},
  \citenamefont {Crowter}, \citenamefont {Dolch}, \citenamefont {Ferdman},
  \citenamefont {Fonseca}, \citenamefont {Gonzalez}, \citenamefont {Jones},
  \citenamefont {Jones}, \citenamefont {Lam}, \citenamefont {Levin},
  \citenamefont {McLaughlin}, \citenamefont {Pennucci}, \citenamefont
  {Stovall},\ and\ \citenamefont {Swiggum}}]{Zhu2015}%
  \BibitemOpen
  \bibfield  {author} {\bibinfo {author} {\bibfnamefont {W.~W.}\ \bibnamefont
  {Zhu}}, \bibinfo {author} {\bibfnamefont {I.~H.}\ \bibnamefont {Stairs}},
  \bibinfo {author} {\bibfnamefont {P.~B.}\ \bibnamefont {Demorest}}, \bibinfo
  {author} {\bibfnamefont {D.~J.}\ \bibnamefont {Nice}}, \bibinfo {author}
  {\bibfnamefont {J.~A.}\ \bibnamefont {Ellis}}, \bibinfo {author}
  {\bibfnamefont {S.~M.}\ \bibnamefont {Ransom}}, \bibinfo {author}
  {\bibfnamefont {Z.}~\bibnamefont {Arzoumanian}}, \bibinfo {author}
  {\bibfnamefont {K.}~\bibnamefont {Crowter}}, \bibinfo {author} {\bibfnamefont
  {T.}~\bibnamefont {Dolch}}, \bibinfo {author} {\bibfnamefont {R.~D.}\
  \bibnamefont {Ferdman}}, \bibinfo {author} {\bibfnamefont {E.}~\bibnamefont
  {Fonseca}}, \bibinfo {author} {\bibfnamefont {M.~E.}\ \bibnamefont
  {Gonzalez}}, \bibinfo {author} {\bibfnamefont {G.}~\bibnamefont {Jones}},
  \bibinfo {author} {\bibfnamefont {M.~L.}\ \bibnamefont {Jones}}, \bibinfo
  {author} {\bibfnamefont {M.~T.}\ \bibnamefont {Lam}}, \bibinfo {author}
  {\bibfnamefont {L.}~\bibnamefont {Levin}}, \bibinfo {author} {\bibfnamefont
  {M.~A.}\ \bibnamefont {McLaughlin}}, \bibinfo {author} {\bibfnamefont
  {T.}~\bibnamefont {Pennucci}}, \bibinfo {author} {\bibfnamefont
  {K.}~\bibnamefont {Stovall}}, \ and\ \bibinfo {author} {\bibfnamefont
  {J.}~\bibnamefont {Swiggum}},\ }\href {\doibase 10.1088/0004-637X/809/1/41}
  {\bibfield  {journal} {\bibinfo  {journal} {Astrophys. J.}\ }\textbf
  {\bibinfo {volume} {809}},\ \bibinfo {pages} {41} (\bibinfo {year}
  {2015})}\BibitemShut {NoStop}%
\bibitem [{\citenamefont {Hofmann}\ and\ \citenamefont
  {M\"{u}ller}(2018)}]{Hofmann2018}%
  \BibitemOpen
  \bibfield  {author} {\bibinfo {author} {\bibfnamefont {F.}~\bibnamefont
  {Hofmann}}\ and\ \bibinfo {author} {\bibfnamefont {J.}~\bibnamefont
  {M\"{u}ller}},\ }\href {\doibase 10.1088/1361-6382/aa8f7a} {\bibfield
  {journal} {\bibinfo  {journal} {Classical Quantum Gravity}\ }\textbf
  {\bibinfo {volume} {35}},\ \bibinfo {pages} {035015} (\bibinfo {year}
  {2018})}\BibitemShut {NoStop}%
\bibitem [{\citenamefont {Koivisto}(2008{\natexlab{b}})}]{Koivisto2008b}%
  \BibitemOpen
  \bibfield  {author} {\bibinfo {author} {\bibfnamefont {T.~S.}\ \bibnamefont
  {Koivisto}},\ }\href {\doibase 10.1103/PhysRevD.78.123505} {\bibfield
  {journal} {\bibinfo  {journal} {Phys. Rev. D}\ }\textbf {\bibinfo {volume}
  {78}},\ \bibinfo {pages} {123505} (\bibinfo {year}
  {2008}{\natexlab{b}})}\BibitemShut {NoStop}%
\bibitem [{\citenamefont {Kehagias}\ and\ \citenamefont
  {Maggiore}(2014)}]{Kehagias2014}%
  \BibitemOpen
  \bibfield  {author} {\bibinfo {author} {\bibfnamefont {A.}~\bibnamefont
  {Kehagias}}\ and\ \bibinfo {author} {\bibfnamefont {M.}~\bibnamefont
  {Maggiore}},\ }\href {\doibase 10.1007/JHEP08(2014)029} {\bibfield  {journal}
  {\bibinfo  {journal} {J. High Energy Phys.}\ } {\bibinfo {volume}
  {8}} (\bibinfo {year} {2014})\ \bibinfo {pages} {29}}\BibitemShut {NoStop}%
\bibitem [{\citenamefont {Conroy}\ \emph {et~al.}(2015)\citenamefont {Conroy},
  \citenamefont {Koivisto}, \citenamefont {Mazumdar},\ and\ \citenamefont
  {Teimouri}}]{Conroy2015}%
  \BibitemOpen
  \bibfield  {author} {\bibinfo {author} {\bibfnamefont {A.}~\bibnamefont
  {Conroy}}, \bibinfo {author} {\bibfnamefont {T.}~\bibnamefont {Koivisto}},
  \bibinfo {author} {\bibfnamefont {A.}~\bibnamefont {Mazumdar}}, \ and\
  \bibinfo {author} {\bibfnamefont {A.}~\bibnamefont {Teimouri}},\ }\href
  {\doibase 10.1088/0264-9381/32/1/015024} {\bibfield  {journal} {\bibinfo
  {journal} {Classical Quantum Gravity}\ }\textbf {\bibinfo {volume} {32}},\
  \bibinfo {pages} {015024} (\bibinfo {year} {2015})}\BibitemShut {NoStop}%
\bibitem [{\citenamefont {Belgacem}\ \emph {et~al.}(2019)\citenamefont
  {Belgacem}, \citenamefont {Finke}, \citenamefont {Frassino},\ and\
  \citenamefont {Maggiore}}]{Belgacem2019}%
  \BibitemOpen
  \bibfield  {author} {\bibinfo {author} {\bibfnamefont {E.}~\bibnamefont
  {Belgacem}}, \bibinfo {author} {\bibfnamefont {A.}~\bibnamefont {Finke}},
  \bibinfo {author} {\bibfnamefont {A.}~\bibnamefont {Frassino}}, \ and\
  \bibinfo {author} {\bibfnamefont {M.}~\bibnamefont {Maggiore}},\ }\href
  {\doibase 10.1088/1475-7516/2019/02/035} {\bibfield  {journal} {\bibinfo
  {journal} {J. Cosmol. Astropart. Phys.}\ } {\bibinfo {volume} {02}} (\bibinfo {year} {2019})\
  \bibinfo {pages} {035}}\BibitemShut {NoStop}%
\bibitem [{\citenamefont {McVittie}(1933)}]{McVittie1933}%
  \BibitemOpen
  \bibfield  {author} {\bibinfo {author} {\bibfnamefont {G.~C.}\ \bibnamefont
  {McVittie}},\ }\href {\doibase 10.1093/mnras/93.5.325} {\bibfield  {journal}
  {\bibinfo  {journal} {Mon. Not. R. Astron. Soc.}\ }\textbf {\bibinfo {volume}
  {93}},\ \bibinfo {pages} {325} (\bibinfo {year} {1933})}\BibitemShut
  {NoStop}%
\bibitem [{\citenamefont {Clifton}\ \emph {et~al.}(2012)\citenamefont
  {Clifton}, \citenamefont {Ferreira}, \citenamefont {Padilla},\ and\
  \citenamefont {Skordis}}]{Clifton2012}%
  \BibitemOpen
  \bibfield  {author} {\bibinfo {author} {\bibfnamefont {T.}~\bibnamefont
  {Clifton}}, \bibinfo {author} {\bibfnamefont {P.~G.}\ \bibnamefont
  {Ferreira}}, \bibinfo {author} {\bibfnamefont {A.}~\bibnamefont {Padilla}}, \
  and\ \bibinfo {author} {\bibfnamefont {C.}~\bibnamefont {Skordis}},\ }\href
  {\doibase 10.1016/j.physrep.2012.01.001} {\bibfield  {journal} {\bibinfo
  {journal} {Phys. Rep.}\ }\textbf {\bibinfo {volume} {513}},\ \bibinfo {pages}
  {1} (\bibinfo {year} {2012})}\BibitemShut {NoStop}%
\bibitem [{\citenamefont {Nesseris}\ and\ \citenamefont
  {Tsujikawa}(2014)}]{Nesseris2014}%
  \BibitemOpen
  \bibfield  {author} {\bibinfo {author} {\bibfnamefont {S.}~\bibnamefont
  {Nesseris}}\ and\ \bibinfo {author} {\bibfnamefont {S.}~\bibnamefont
  {Tsujikawa}},\ }\href {\doibase 10.1103/PhysRevD.90.024070} {\bibfield
  {journal} {\bibinfo  {journal} {Phys. Rev. D}\ }\textbf {\bibinfo {volume}
  {90}},\ \bibinfo {pages} {024070} (\bibinfo {year} {2014})}\BibitemShut
  {NoStop}%
\end{thebibliography}
%

\end{document}